\documentclass[aps,prl,preprint,tightenlines,superscriptaddress,showpacs,byrevtex]{revtex4}

\usepackage{graphics}
\usepackage{dcolumn}  
\usepackage{color}
\usepackage{epsf}
\usepackage{latexsym}
\usepackage{amssymb}
\graphicspath{{ps}}

\def\TME{\tau^-\rightarrow\mu^-\eta}
\def\TMEP{\tau^-\rightarrow\mu^-\eta'}
\def\TEE{\tau^-\rightarrow e^- \eta}
\def\TEEP{\tau^-\rightarrow e^- \eta'}
\def\TMP{\tau^-\rightarrow\mu^-\pi^0}
\def\TEP{\tau^-\rightarrow e^- \pi^0}
\def\T3M{\tau^-\rightarrow 3\mu}
\def\EGG{\eta\rightarrow \gamma\gamma}
\def\E3P{\eta\rightarrow \pi^+\pi^-\pi^0}
\def\EP{\eta'\rightarrow \pi^+\pi^-\eta}
\def\TT{\tau\tau}
\def\TTC{\tau^+\tau^-}

\def\GG{\gamma\gamma}

\def\EGG{\eta\rightarrow \gamma\gamma}
\def\EIP3{\eta\rightarrow 3\pi}
\def\MTME{\mu^-\eta}
\def\MTMEP{\mu^-\eta'}
\def\MTEE{e^- \eta}
\def\MTEEP{e^- \eta'}
\def\MTMP{\mu^-\pi^0}
\def\MTEP{e^- \pi^0}

\def\Minv{M_{\ell^- M^0}}
\def\DelE{\Delta E}

\def\RA{\rightarrow}
\def\10E7{\times 10^7}
\def\lrat{\mathcal{P}}
\def\like{\mathcal{L}}
\begin{document}

\preprint{\tighten\vbox{\hbox{\hfil Belle Preprint 2005-10}
                        \hbox{\hfil KEK   Preprint 2004-107}
                        \hbox{\hfil Intended for {\it PLB}}
}}

\vspace*{15mm}

\title{\quad\\[0.5cm] 
Search for lepton flavor violating
decays $\tau^- \rightarrow$
$\ell^- \pi^0, \ell^- \eta, \ell^- \eta'$ 
}%

\begin{abstract}

We have searched for lepton flavor violating
semileptonic
$\tau^-$ decays 
using
a data sample of 153.8 fb$^{-1}$  accumulated with the Belle detector 
at the KEKB $e^+e^-$ collider. 
For
the
six decay modes studied, 
the observed yield is compatible with the estimated background
and the following upper limits are set at the 90\% confidence level:
$\mathcal{B}(\TEE ) <  2.3   \times 10^{-7}$, 
$\mathcal{B}(\TME ) <  1.5   \times 10^{-7}$, 
$\mathcal{B}(\TEP ) <  1.9   \times 10^{-7}$, 
$\mathcal{B}(\TMP ) <  4.1   \times 10^{-7}$, 
$\mathcal{B}(\TEEP) <  10    \times 10^{-7}$, 
and
$\mathcal{B}(\TMEP) <  4.7   \times 10^{-7}$.
These results are 10 to 64 times more
restrictive
than
previous limits.
\end{abstract}

%
%

\affiliation{Budker Institute of Nuclear Physics, Novosibirsk}
\affiliation{Chiba University, Chiba}
\affiliation{Chonnam National University, Kwangju}
\affiliation{University of Cincinnati, Cincinnati, Ohio 45221}
\affiliation{University of Hawaii, Honolulu, Hawaii 96822}
\affiliation{High Energy Accelerator Research Organization (KEK), Tsukuba}
\affiliation{Hiroshima Institute of Technology, Hiroshima}
\affiliation{Institute of High Energy Physics, Chinese Academy of Sciences, Beijing}
\affiliation{Institute of High Energy Physics, Vienna}
\affiliation{Institute for Theoretical and Experimental Physics, Moscow}
\affiliation{J. Stefan Institute, Ljubljana}
\affiliation{Kanagawa University, Yokohama}
\affiliation{Korea University, Seoul}
\affiliation{Kyungpook National University, Taegu}
\affiliation{Swiss Federal Institute of Technology of Lausanne, EPFL, Lausanne}
\affiliation{University of Ljubljana, Ljubljana}
\affiliation{University of Maribor, Maribor}
\affiliation{University of Melbourne, Victoria}
\affiliation{Nagoya University, Nagoya}
\affiliation{Nara Women's University, Nara}
\affiliation{National Central University, Chung-li}
\affiliation{National United University, Miao Li}
\affiliation{Department of Physics, National Taiwan University, Taipei}
\affiliation{H. Niewodniczanski Institute of Nuclear Physics, Krakow}
\affiliation{Nihon Dental College, Niigata}
\affiliation{Niigata University, Niigata}
\affiliation{Osaka City University, Osaka}
\affiliation{Osaka University, Osaka}
\affiliation{Panjab University, Chandigarh}
\affiliation{Peking University, Beijing}
\affiliation{Princeton University, Princeton, New Jersey 08544}
\affiliation{University of Science and Technology of China, Hefei}
\affiliation{Seoul National University, Seoul}
\affiliation{Sungkyunkwan University, Suwon}
\affiliation{University of Sydney, Sydney NSW}
\affiliation{Tata Institute of Fundamental Research, Bombay}
\affiliation{Toho University, Funabashi}
\affiliation{Tohoku Gakuin University, Tagajo}
\affiliation{Tohoku University, Sendai}
\affiliation{Department of Physics, University of Tokyo, Tokyo}
\affiliation{Tokyo Institute of Technology, Tokyo}
\affiliation{Tokyo Metropolitan University, Tokyo}
\affiliation{Tokyo University of Agriculture and Technology, Tokyo}
\affiliation{University of Tsukuba, Tsukuba}
\affiliation{Virginia Polytechnic Institute and State University, Blacksburg, Virginia 24061}
\affiliation{Yonsei University, Seoul}
   \author{Y.~Enari}\affiliation{Nagoya University, Nagoya} 
   \author{N.~Sato}\affiliation{Nagoya University, Nagoya} 
   \author{K.~Abe}\affiliation{High Energy Accelerator Research Organization (KEK), Tsukuba} 
   \author{K.~Abe}\affiliation{Tohoku Gakuin University, Tagajo} 
   \author{H.~Aihara}\affiliation{Department of Physics, University of Tokyo, Tokyo} 
   \author{Y.~Asano}\affiliation{University of Tsukuba, Tsukuba} 
   \author{V.~Aulchenko}\affiliation{Budker Institute of Nuclear Physics, Novosibirsk} 
   \author{S.~Bahinipati}\affiliation{University of Cincinnati, Cincinnati, Ohio 45221} 
   \author{A.~M.~Bakich}\affiliation{University of Sydney, Sydney NSW} 
   \author{I.~Bedny}\affiliation{Budker Institute of Nuclear Physics, Novosibirsk} 
% ...
\collaboration{The Belle Collaboration}

\maketitle

\section{Introduction}
Processes with Lepton Flavor Violation (LFV)
provide some of the most promising searches
for physics beyond the Standard Model (SM). In the charged lepton
sector LFV is usually considered for purely leptonic processes 
such as radiative decays
of $\mu$ and $\tau$~\cite{hisano}
or their decays into
three charged leptons~\cite{higgs_mediated}. 
The semileptonic $\tau$ decays
$\tau^- \to \ell^- M^0$
(where $\ell^- = e^-, \mu^-$ and $M^0 = \pi^0, \eta, \eta'$)
provide
another good source of information about possible LFV mechanisms from
supersymmetry (SUSY) motivated models~\cite{Sher,Brignole:2004ah} to models
with heavy Dirac neutrinos~\cite{gonzalez,ilakovac};
they also allow more general bounds to be placed on
the scale of new physics~\cite{black}. 
Experiments at B factories, 
where $\tau$ lepton pairs are copiously produced, 
have
substantially
increased the sensitivity to LFV decays, bringing it close
to various theoretical expectations. 
Recently we performed a search
for the decay $\tau^- \to \mu^-\eta$ based on a data sample of
84.3 fb$^{-1}$ and placed a new upper limit on its branching fraction,
$\mathcal{B}(\tau^- \to \mu^-\eta) < 3.4 \times 10^{-7}$~\cite{res_belle},
28 times better than the previous limit from CLEO~\cite{cleo} and comparable
to predictions for some SUSY models~\cite{Sher,Brignole:2004ah}.

We present here an updated result for $\TME$
and new searches for
the decay modes 
$\tau^- \rightarrow \mu^-\pi^0$, $\tau^- \rightarrow \mu^-\eta'$,
$\tau^- \rightarrow e^-\pi^0$, $\tau^- \rightarrow e^-\eta$,
and $\tau^- \rightarrow e^-\eta'$, 
based on a data sample of 153.8 fb$^{-1}$,
equivalent to 137.2 $\times 10^6$ $\TTC$ pairs, 
collected at the $\Upsilon(4S)$ resonance
with the Belle detector
at the KEKB energy-asymmetric $e^+e^-$ collider~\cite{KEKB}. 
Due to
this
much higher integrated luminosity
our sensitivity to the
decay modes
$\tau^- \rightarrow \ell^- \pi^0$ and $\ell^- \eta$
is considerably better
than in previous searches with MARK II~\cite{mark}, Crystal Ball~\cite{ball},
ARGUS~\cite{argus} and CLEO~\cite{cleo}. The decay modes
$\tau^- \rightarrow \ell^- \eta'$
are studied here for the first time.
Unless otherwise stated, 
charge conjugate
modes
are implied
throughout the paper.

The Belle detector is a large-solid-angle magnetic spectrometer that
consists of a silicon vertex detector (SVD), 
a 50-layer central drift chamber (CDC), 
an array of aerogel threshold \v{C}erenkov counters (ACC), 
a barrel-like arrangement of time-of-flight scintillation counters (TOF), 
and
an electromagnetic calorimeter (ECL) comprised of CsI(Tl) crystals
located inside a superconducting solenoid coil
that provides a 1.5~T magnetic field.  
An iron flux-return located outside of the coil is instrumented
to detect $K_L^0$ mesons and to identify muons (KLM).  
A more detailed description of the detector can be found in Ref.~\cite{belle}. 

For Monte Carlo (MC) studies, the following programs have been used 
to generate
background
events:
KORALB/TAUOLA~\cite{tauola} for $\TTC$ processes, 
QQ~\cite{qq} for $B\bar{B}$ and continuum, 
BHLUMI~\cite{bhabha} for Bhabha, KKMC~\cite{kkmc} for $\mu^+\mu^-$ and 
AAFH~\cite{aafhb} for two-photon processes.
We use the following sizes of MC
samples:
428 fb$^{-1}$ of generic $\TTC$, 
232 fb$^{-1}$ of $q\bar{q}$ with $q = u,d,s$,
and 160 fb$^{-1}$ with $q=c$, 
90  fb$^{-1}$ of $B\bar{B}$, 
119 fb$^{-1}$ of $\mu^+\mu^-$, 
9.1 fb$^{-1}$ of Bhabha
and
205 fb$^{-1}$ of
two-photon processes.
Signal MC is generated by
KORALB/TAUOLA. To take a range of possible
$\tau^- \rightarrow \ell^- M^0$ production mechanisms into account, 
we generate events under three different
assumptions: uniform angular distribution in the $\tau$ rest frame,
$V-A$ interactions, and $V+A$ interactions.
We initially assume all $\tau$ decays 
to have a uniform angular distribution
and take the other
hypotheses
into account later.
The Belle detector response is simulated by a GEANT3~\cite{geant} 
based program.

\section{Data Analysis}

We follow
the same
principles of event selection
as those in the 
$\tau^-\rightarrow\mu^-\eta$ search~\cite{res_belle}.
Kinematical variables with a CM superscript are
calculated in the center-of-mass frame, 
and other variables are calculated in the laboratory frame
unless otherwise stated.
We look for an event with the following particles: 
\begin{displaymath}
\begin{array}{c}
 \left\{\tau^-\rightarrow \ell^- + M^0 \right\}
 + \left\{\tau^+\rightarrow ({\rm track})^+ + n \gamma + X \right\}, 
\end{array}
\end{displaymath}
where the system $X$ is unobserved.
In other words, the events should be consistent with a $\TTC$ event,
in which the $\tau^-$ decays into
a lepton $\ell^-$
and a pseudoscalar meson $M^0$ (signal side)
and the $\tau^+$ decays into a charged track,
any number $n \geq 0$ of photons, and one or more neutrinos (tag side).
The charged track on the tag side should not be an electron (muon) 
if the lepton on the signal side is an electron (muon), in 
order to avoid contamination by Bhabha ($\mu^+\mu^-$) events. 
We reconstruct mesons in the following modes: $\pi^0 \to \gamma \gamma$,
$\eta \to \gamma \gamma$,  $\eta \to \pi^+\pi^-\pi^0$, 
$\eta^{\prime} \to \pi^+\pi^-\eta, \eta \to \gamma \gamma$.
When pseudoscalar mesons $M^0$ are reconstructed 
from $\gamma\gamma$, the event has a 1-1 prong configuration. 
When the $\eta$ is reconstructed from $\pi^+\pi^-\pi^0$,
or the $\eta'$ is reconstructed in $\eta' \RA \pi^+\pi^-\eta$, $\EGG$,  
the event has a 1-3 prong configuration. 
The signal side thus contains at least two photons in all cases.

For the 1-1 prong selection, 
the candidate should contain exactly two 
oppositely charged tracks and 
two or more photons, two of which form a $\pi^0$ or an $\eta$. 
We require the momentum of each track, $p$, and the energy of each
photon, $E_{\gamma}$, to satisfy
$p > 0.1$ GeV/c and $E_{\gamma} > 0.1$ GeV. 
The tracks and photons are required to be
detected within the CDC acceptance, 
$-0.866$ $< \cos\theta < 0.956$.
In order to exclude
Bhabha, $\mu^+\mu^-$ and two-photon events, 
which otherwise contribute a large background, 
the total
visible
energy in the CM frame is required to satisfy 
5~GeV~$< E^{\rm{CM}}_{\rm{total}} < $~10~GeV. 
The event is subdivided into two hemispheres by a plane perpendicular to 
the thrust axis
evaluated from all tracks and
photons
satisfying the above requirements. 

The lepton flavor is identified based on likelihood ratios
calculated from the response of various subsystems of the detector.
For electron identification, 
the likelihood ratio is defined as
$\lrat_e = \like_e / (\like_e + \like_{x})$, 
where $\mathcal{L}_e$ and  $\mathcal{L}_x$ are the likelihoods
for electron and other flavor hypotheses, respectively, 
determined using
the matching between the position
of the charged track trajectory and the cluster position in the ECL, 
the ratio of the shower energy
in the ECL to the momentum measured by the CDC,
the shower shape of the cluster in the ECL, 
specific ionisation ($dE/dx$)
in the CDC
and the light yield in the ACC~\cite{eid}.
For muon identification, the likelihood ratio is
$\lrat_{\mu} = \like_{\mu} / (\like_\mu + \like_K+ \like_\pi)$, 
where ${\mathcal{L}}_\mu$, ${\mathcal{L}}_\pi$, and ${\mathcal{L}}_K$ are
the likelihoods for the muon, pion and kaon
hypotheses respectively,
based on the matching quality and
penetration depth of associated hits in the KLM~\cite{muid}.
For electron candidates $\lrat_{e} > $ 0.9 is required,
identifying electrons with an efficiency $\eta_{e} = (92.4 \pm 0.4)$\%;
the pion fake rate
(the probability that a pion is misidentified as an electron) 
$\kappa_{e} = (0.25 \pm 0.02)$\%.
For muons  $\eta_{\mu} = (87.5 \pm 0.3)$\% 
and $\kappa_{\mu} = (1.4 \pm 0.1)$\% are obtained.
If the track satisfies
${\mathcal{P}}_{e}>0.6$ or ${\mathcal{P}}_{\mu}>0.6$ on the tag side, 
i.e. is lepton-like,
the constraints
$n_{\gamma}^{\rm tag}\leq$ 2
and 
$p_{e/\mu}^{\rm tag}$ 0.7 GeV/$c$
are applied
to suppress events including photons
from initial state radiation or beam background.
Otherwise, no constraints are applied on the tag side.
The lepton flavor identification on the signal side is
postponed until the last stage of selection.

\begin{figure}[th]
 \resizebox{.35\textwidth}{0.3\textwidth}{\includegraphics{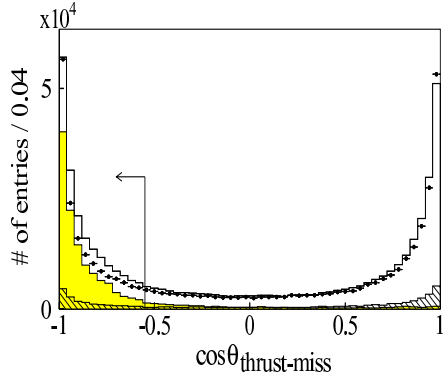}}
\caption{\small
 $\cos\theta^{\rm CM}_{\rm thrust-miss}$ distribution
 for the $\tau^- \rightarrow \mu^-\eta$ decay mode.
The data (points with error bars), 
total background MC (open histogram), 
non-$\tau^+\tau^-$ background MC (hatched histogram), 
and signal MC (shaded histogram) are shown.
 The selected region is indicated by the arrow.
}
\label{fig:cos_thrust-miss}
\end{figure}

\begin{table}[b]
 \caption{The resolutions of $\Minv$ and $\Delta E$ for each mode.
 The ``low'' or ``high'' superscript indicates the lower or higher energy side 
of the peak; 
$\sigma_{\Minv}$ are in MeV/$c^2$ and $\sigma_{\DelE}$ are in MeV.
 }
\label{tab:resolution}
 \begin{tabular}{|l|l||c|c|c|c|} \hline \hline
 Mode   & Subdecay mode & $\sigma_{\Minv}^{\rm low}$ & $\sigma_{\Minv}^{\rm high}$ & ~~$\sigma_{\DelE}^{\rm low}$ ~~& ~~$\sigma_{\DelE}^{\rm high}$~~ \\  \hline
 $\TEE$ & $\EGG$                           &25.8 & 15.3 & 79.3 & 34.7 \\
        & $\E3P$ ~($\pi^0 \rightarrow \gamma\gamma$) &17.5 &  9.7 & 44.2 & 25.5 \\ 
 $\TME$ & $\EGG$                           &25.7 & 14.2 & 57.1 & 35.2 \\ 
        & $\E3P$ ~($\pi^0 \rightarrow \gamma\gamma$) &13.4 &  9.6 & 43.2 & 22.1 \\ 
 $\TEP$ & $\pi^0 \rightarrow \gamma\gamma$ &25.7 & 14.7 & 69.5 & 35.2 \\ 
 $\TMP$ & $\pi^0 \rightarrow \gamma\gamma$ &23.6 & 14.5 & 69.8 & 38.9  \\ 
 $\TEEP$& $\EP$ ~($\eta \rightarrow \gamma\gamma$) &22.0 & 14.0 & 62.7 & 26.7 \\ 
 $\TMEP$& $\EP$ ~($\eta \rightarrow \gamma\gamma$) &18.3 & 11.7 & 55.6 & 25.9 \\ 
  \hline \hline
 \end{tabular} 
\end{table}

The momentum of a $\pi^0$ or $\eta$ meson produced in a 
two-body $\tau$-decay is on average 
higher than that of $\pi^0/\eta$ mesons from other sources, 
so that photons from $\pi^0$ or $\eta$ are required to 
have a rather high energy $E_{\gamma}>$ 0.30 (0.22) GeV in 
the case of $\ell^-$ = $e^-$($\mu^-$). 
To further reduce the backgrounds, the cosine of the opening angle 
between $\ell^-$ and $\gamma\gamma$ 
on the signal side must satisfy 
$0.5 < \cos\theta_{\hbox{\scriptsize $\ell^-$-$\gamma\gamma$}} < 0.95$. 
When reconstructing an $\eta$ meson candidate,
a $\pi^0$ veto is applied: 
defining the resolution-normalized $\pi^0$ mass
$S^{\pi^0}_{\gamma\gamma'}=(m_{\gamma\gamma'}-0.135~\mbox{GeV/}c^2)/
\sigma^{\pi^0}_{\gamma\gamma'}$, we reject events
where a signal-side photon $\gamma$ and a tag-side photon $\gamma'$
satisfy $-5<S^{\pi^0}_{\gamma\gamma'}<+5$;
the resolution
$\sigma_{\gamma \gamma'}^{\pi^0}$
is in the range 5--8 MeV/$c^2$.
This $\pi^0$ veto rejects 86\% of the BG events while retaining 75\% 
of the signal. 
To ensure that the missing particles are neutrinos rather than 
$\gamma$'s or charged particles 
that fall outside the detector acceptance, we require that the 
direction of the missing momentum satisfy the condition
$-0.866< \cos\theta_{\rm{miss}}<0.956$.
Since neutrinos are emitted only on the tag side, the direction 
of the missing momentum $p_{\rm{miss}}$ 
should be contained on the tag side. 
We define the angle 
$\theta^{\rm CM}_{\rm thrust-miss}$
between the thrust axis of the event (in the signal hemisphere)
and the missing momentum vector, 
and require
$\cos\theta_{\rm{thrust-miss}}^{\rm CM}<-0.55$
as shown in Fig.~\ref{fig:cos_thrust-miss}. 
At this level of selection the dominant
background
is that of $\tau^+\tau^-$
events (95\%) whereas $q\bar{q}$ events and other sources
constitute
only 5\%.
The correlation between the missing momentum, $p_{\rm{miss}}$, 
and the missing mass squared, $m_{\rm{miss}}^2$, is 
used to remove the generic $\tau^+\tau^-$ and $q\overline{q}$ continuum 
contributions: 
$p_{\rm{miss}} > -2.616 m^2_{\rm{miss}}-0.191$ and $p_{\rm{miss}} > 1.0 m^2_{\rm{miss}}-1$,
 where $p_{\rm miss}$ is in GeV/$c$ and $m_{\rm miss}$ is in GeV/$c^2$.

The yield of signal events is finally 
determined in the $\Minv$--$\Delta E$ plane, 
where $\Minv$ is the invariant mass of the $\ell^- M^0$ system and 
$\Delta E$ is the difference between the energy of the $\ell^- M^0$ 
system and the beam energy in the CM frame. 
When deciding on our selection criteria,
we blinded the signal region
$M_\tau - 5\sigma^{\rm low}_{\Minv}< \Minv < M_\tau + 5\sigma^{\rm high}_{\Minv}$
and 
$-0.5$ GeV $< \Delta E < 0.5$ GeV.
We define
sideband regions of $\Minv$ and $\Delta E$
as
$M_\tau - 10\sigma^{\rm low}_{\Minv}< \Minv < M_\tau - 5\sigma^{\rm high}_{\Minv}$,
$M_\tau + 5\sigma^{\rm low}_{\Minv}< \Minv < M_\tau + 10\sigma^{\rm high}_{\Minv}$
and 
$-10\sigma_{\Delta E}^{\rm low} < \Delta E < -0.5$ GeV, 
0.5 GeV $< \Delta E < 10\sigma_{\Delta E}^{\rm high}$, respectively.
The resolutions, 
$\sigma_{\Minv}^{\rm low/high}$
and 
$\sigma_{\Delta E}^{\rm low/high}$, 
of $\Minv$ and $\Delta E$, 
are evaluated from the distributions of signal MC
around the peak using an asymmetric Gaussian shape 
to account for initial state radiation and ECL energy leakage for photons.
In these expressions, $M_\tau$ is the central value of the
reconstructed $\tau$ mass for signal, evaluated in MC:
it is consistent with the known $\tau$ mass within
1.2
-- 13.0 MeV for each mode.
The resulting resolutions are summarized in Table \ref{tab:resolution}. 

\begin{figure}[t]
 \centerline{
 \resizebox{.35\textwidth}{0.3\textwidth}{\includegraphics{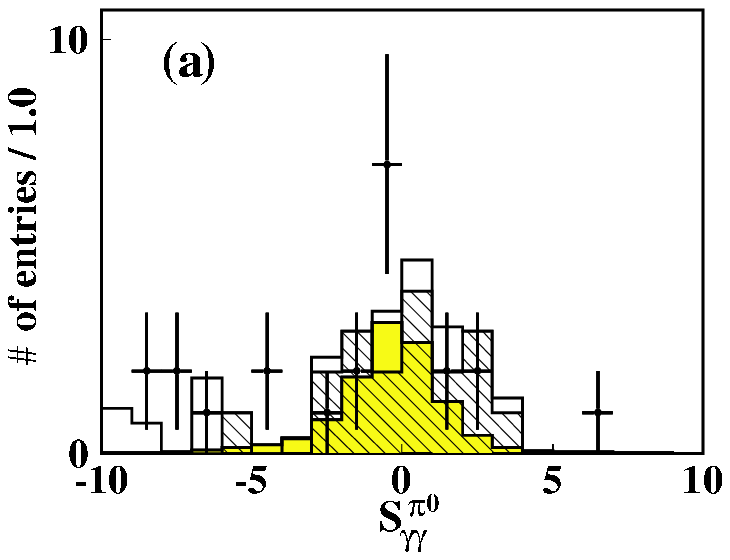}}
 \resizebox{.35\textwidth}{0.305\textwidth}{\includegraphics{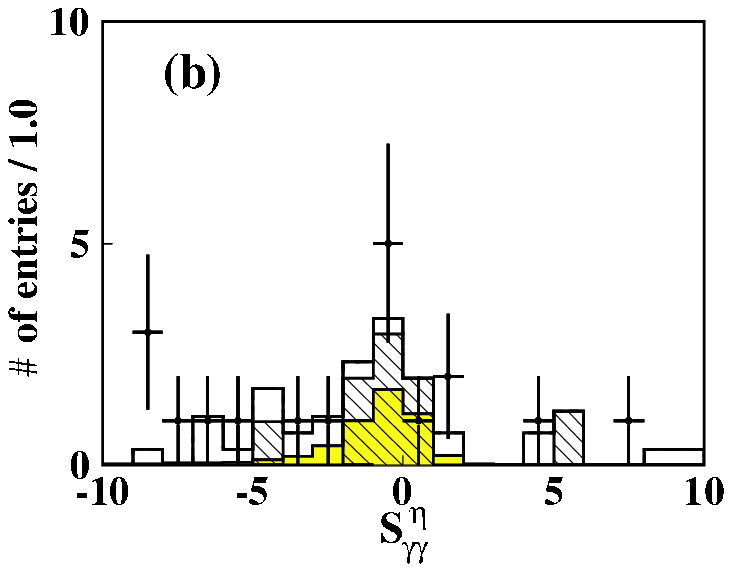}}
 \resizebox{.35\textwidth}{0.3\textwidth}{\includegraphics{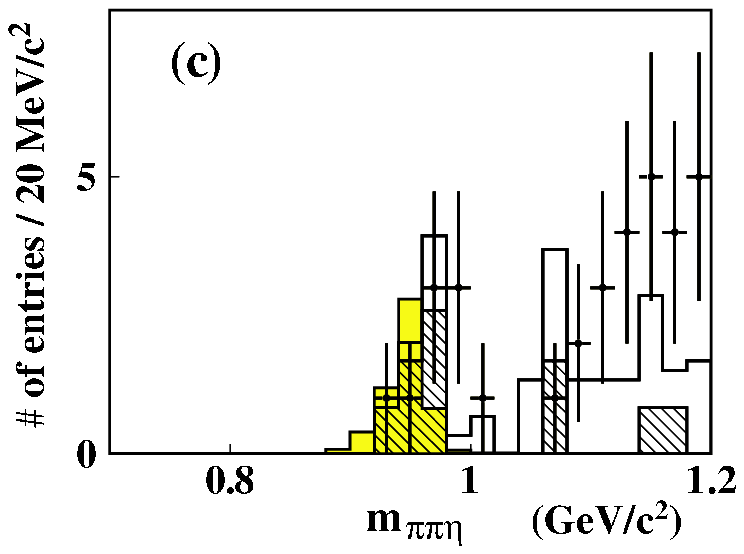}}
}
\caption{\small
$S_{\gamma\gamma}^{\pi^0}$, $S_{\gamma\gamma}^{\eta}$ and $m_{\pi^+\pi^-\eta}$
distributions for
(a) $\tau^- \rightarrow \mu^-\pi^0$, 
(b) $\tau^- \rightarrow \mu^-\eta$ 
and (c)
$\tau^- \rightarrow \mu^-\eta'$
decay modes, respectively,  
prior to final cuts on the signal side
(see the text).
The data (points with error bars), 
total background MC (open histogram), 
non-$\tau^+\tau^-$ background MC (hatched histogram), 
and signal MC
(shaded histogram, corresponding to the branching fraction of 5$\times$10$^{-7}$)
are shown.
}
\label{fig:pi0_eta_etap_for_muon}
\end{figure}

The resulting numbers of data and MC events
in the $\Minv$--$\Delta E$ sideband region
after applying 
a loose mass
requirement
for $\pi^0/\eta$, 
$-5 < S_{\gamma\gamma}^{\pi^0/\eta} < 5$, 
are summarized 
in Table \ref{tab:number_of_ev_at_Lv2}.
The $S_{\gamma\gamma}^{\pi^0}$ and $S_{\gamma\gamma}^{\eta}$
distributions
for the modes with a final state muon
are shown in Fig.\ref{fig:pi0_eta_etap_for_muon} (a) and (b), 
respectively.
Within the statistical uncertainty
the sideband data and
background
MC yields are
consistent with each other in all four modes.

At the last stage of the selection, 
the $\pi^0$ or $\eta$ candidate is required to satisfy 
$-5 < S^{\pi^0/\eta}_{\gamma\gamma} < 3$, 
where $S^{\pi^0}_{\gamma\gamma}$ was defined above, 
and 
$S^{\eta}_{\gamma\gamma}=(m_{\gamma\gamma}-0.547~\mbox{GeV/}c^2)/\sigma^{\eta}_{\gamma\gamma}$ 
with $\sigma^{\eta}_{\GG}$ = 11--13 MeV$/c^2$. 
The conditions
${\mathcal{P}}_{e/\mu} > 0.9$ and $p > 0.7$ GeV/c are imposed
on
lepton tracks on the signal side.
In addition, ${\mathcal{P}}_{e/\mu} < 0.6$ is required on the tag side 
for rejection of Bhabha or $\mu^+\mu^-$ events.

\begin{table}[t]
 \caption{Numbers of remaining events for each mode
at the same selection stage as in Fig.~\ref{fig:pi0_eta_etap_for_muon}, 
prior to final cuts on the signal side (see the text).
}
 \label{tab:number_of_ev_at_Lv2}
 \begin{tabular}{|l|l||c|c|} \hline \hline
 Mode                            & Subdecay mode                      & $N_{\rm side}^{\rm DATA}$  & $N_{\rm side}^{\rm MC}$ \\ \hline
 $\tau^- \rightarrow e^-\eta$    & $\eta \rightarrow \gamma\gamma$    & 3                  & 1.47  $\pm$ 0.73 \\
                                 & $\eta \rightarrow \pi^+\pi^-\pi^0$ & 5                  & 0.34  $\pm$ 0.11 \\
 $\tau^- \rightarrow \mu^-\eta$  & $\eta \rightarrow \gamma\gamma$    & 17                 & 14.4  $\pm$ 3.2 \\
                                 & $\eta \rightarrow \pi^+\pi^-\pi^0$ & 7                  & 5.2  $\pm$ 3.6 \\
 $\tau^- \rightarrow e^-\pi^0$   & $\pi^0 \rightarrow \gamma\gamma$   & 2                  & 2.2   $\pm$ 0.9 \\
 $\tau^- \rightarrow \mu^-\pi^0$ & $\pi^0 \rightarrow \gamma\gamma$   & 22                 & 25.8  $\pm$ 4.5 \\
 $\tau^- \rightarrow e^-\eta'$   & $\eta' \rightarrow \pi^+\pi^-\eta$ & 12                 & 11.6  $\pm$ 2.3 \\
 $\tau^- \rightarrow \mu^-\eta'$ & $\eta' \rightarrow \pi^+\pi^-\eta$ & 33                 & 22.9  $\pm$ 3.5 \\ 
 \hline \hline
 \end{tabular}
\end{table}

For $\E3P$ and $\eta'\RA\pi^+\pi^-\eta$ decays with 
$\eta\RA\gamma\gamma$, 
i.e. 1-3 prong modes, 
we search for events containing four charged tracks 
(net charge = 0) and two or more photons. 
Because of the higher multiplicity compared to the 1-1 prong modes, 
the detection efficiency becomes smaller. 
On the other hand, the additional reconstruction constraint in 
the $\eta / \eta'$ decay chain 
improves the background rejection power.
The selection criteria are similar to those in the 
$\eta \to \gamma\gamma$ case with the differences listed below.

The minimum photon energy $E_{\gamma}$ requirement
is reduced from 0.1 GeV to 0.05 GeV,
and
the tighter 
conditions
for photons from $\pi^0/\eta$ are also removed,
since the photons from these decay modes have a softer 
energy distribution compared to that in the 1-1 prong case.
 At least three tracks and two or more photons are required
 in the signal hemisphere. 
One track must be identified as an electron or a muon (${\mathcal{P}}_{{e}/\mu}>$ 0.9),
but particle identification is
not performed on the other two tracks, 
at first they are treated
as pions.
We also require that one $\pi^0$ is reconstructed with
$-5 < S_{\GG}^{\pi^0} < 5$ from 
the photons in the signal hemisphere. 
Without any PID for pion candidate tracks on the signal side, 
possible photon conversions can result in a fake event.
In order to remove such a contribution, 
we examine 
     the effective mass-squared $M^2_{ee}$ of any two tracks 
(to which the electron mass is assigned) 
     in the signal hemisphere. 
     A sharp peak at $M^2_{ee} < 0.01$ 
     $(\rm{GeV}/c^2)^2$ is seen in the Bhabha MC but not in the signal and 
     generic $\tau^+\tau^-$ MC's, and data also exhibit a tiny peak at 
$M^2_{ee}\sim 0$ $(\rm{GeV}/c^2)^2$.
To avoid a large reduction of the signal detection efficiency, 
\textcolor{black}{
we impose a requirement on PID in addition to 
that on $M^2_{ee}$:
events are rejected 
which have $M^2_{ee} < 0.01$ GeV/$c^2$
and ${\mathcal{P}}_{e} > 0.6$
}%
for one of the two tracks. 
     As a result, 60\% of the Bhabha originated events are rejected 
     while the signal efficiency is reduced by 2.3\% only.
In addition, the following constraints are also required:
     $p_{\rm{miss}} > -1.5 m^2_{\rm{miss}}-1.0$ and $p_{\rm{miss}} > 0.8 m^2_{\rm{miss}}-0.2$,
 where $p_{\rm miss}$ is in GeV/$c$ and $m_{\rm miss}$ is in GeV/$c^2$.

The $\pi^0$ and $\eta$ from $\gamma\gamma$  are extracted by imposing 
the conditions $-3 < S_{\gamma\gamma}^{\pi^0} < 3$  and 
$-5 < S_{\gamma\gamma}^{\eta} < 3$, respectively.  
The requirements ${\mathcal{P}}_{e/\mu} > 0.9$ and $p > 0.7$ GeV/c are imposed 
on the tracks on the signal side as in the 1-1 prong case.  
\textcolor{black}{
After applying the above conditions as well as 
the requirements on the invariant mass
1.3 GeV/$c^2$ $< \Minv <$ 2.0 GeV/$c^2$, 
energy difference
$-1.0$ GeV $< \Delta E <$ 0.5 GeV
and a loose requirement on $\eta$ ($\eta'$) mass
$m_{\pi^+\pi^-\pi^0} < 0.65$ GeV/$c^2$ 
($m_{\pi^+\pi^-\eta} < 1.2$ GeV/$c^2$),  
}%
\textcolor{black}{
we obtain the resulting number of data and MC events
summarized 
}%
in Table \ref{tab:number_of_ev_at_Lv2}. 
The $m_{\pi^+\pi^-\eta}$ distribution of
the mode with a final state muon
after these
requirements
is shown in Fig.~\ref{fig:pi0_eta_etap_for_muon} (c).

Again, agreement between sideband data and
background
MC is observed within the 
statistical uncertainty, 
except for 
the decay mode $\tau^- \rightarrow e^- \eta$, $\eta \rightarrow \pi^+\pi^-\pi^0$.
Since in this mode
PID has better rejection power of hadronic background
than in the modes with a final state muon, 
the $\eta$ candidates should be mostly fake ones. 
In fact, 
four of the remaining five data events
are in the $\eta$ mass sideband regions
and thus will be rejected 
at the last stage by
the tighter $\eta$ and $\eta'$ mass requirement
0.5260 GeV/$c^2 < m_{\pi^+\pi^-\pi^0} <$ 0.5656 GeV/$c^2$ 
and
0.9027 GeV/$c^2 < m_{\pi^+\pi^-\eta} <$ 0.9798 GeV/$c^2$,
corresponding to the $\pm 3 \sigma$ regions, 
where the resolutions are estimated from signal MC.

\section{BG Estimations and Branching Fractions}

The signal detection efficiency for each mode is evaluated 
by using signal MC.
We take an elliptically shaped signal region in the $\Minv$--$\DelE$ plane 
with a signal acceptance, $\Omega$, of 90\%, 
which is more sensitive than a box shaped signal region.   
To obtain the highest sensitivity,
all geometrical parameters of the ellipse,
such as the position of the center, the orientation, 
and the length of the major and minor axes,
are determined to minimize its area
with the constraint on the acceptance to be 90\%.

As explained in the previous section, 
an analysis of the
background
MC distributions shows that 
they agree well with data in the sideband regions. 
We also confirmed that no peaking
background
structure,  
which mimics the signal, is observed in the signal region.  

The expected number of background events ($b_0$) in the signal 
elliptical region is estimated from sideband data events as follows:
for
the decay modes 
$\tau^- \rightarrow \mu^-\eta~(\eta \rightarrow \gamma\gamma)$, 
$\tau^- \rightarrow \mu^-\pi^0$ 
and 
$\tau^- \rightarrow \mu^-\eta'$, 
we assume that the background distribution is described
by a polynomial function in $M_{\ell^- M^0}$
and
a Gaussian function in $\Delta E$ over the $\pm 10 \sigma$ region. 
We determine the shape of the function
by fitting to the MC events including
the blinded region with its normalization from the data sideband.
Moreover, 
taking into account the uncertainties
from assuming the background distribution,
we evaluate $b_0$ for these modes by another method:
we assume a flat distribution
over the $\pm 10 \sigma$ region of $M_{\ell^-M^0}$
and the regions 
$-0.4$ GeV $< \Delta E < 0.20$ GeV,
$-0.5$ GeV $< \Delta E < 0.33$ GeV
and
$-0.4$ GeV $< \Delta E < 0.20$ GeV
for the $\mu^- \eta~(\eta \rightarrow \gamma\gamma)$, 
$\mu^- \pi^0$, and $\mu^- \eta'$ modes, respectively.
The differences between the values
evaluated by the first and the second method,
which are 21\%, 10\% and 49\%, respectively, 
are
treated as an additional uncertainty.
For the other five decay modes, 
we evaluate $b_0$ from sideband data events
by simply assuming a flat distribution
over $\pm 10 \sigma$ region of $M_{\ell^- M^0}$--$\Delta E$ plane,
since the remaining number of events
is
too small 
to determine the background shape by fitting to those events.
The uncertainty on $b_0$ is quite
large ($\leq$ 100\%), particularly in the electron modes, because the 
number of events remaining in the sideband region is very small. 
For $\TEP$, there are no events in the sidebands, 
and we set 
$b_0 = 0.0^{+0.4}_{-0.0}$ where the error is calculated by assuming 2.44 events
in the sideband. The resultant $b_0$ are listed in Table \ref{tab_af_open}.

\begin{table}[b]
\caption{\small Results of the final event selection for the 
individual modes:
$\epsilon$ is the detection efficiency, 
$\mathcal{B}_{M^0}$ is the branching fraction for the $M^0$ decay, 
$N^{\rm DATA}_{\rm side}$ and $N^{\rm MC}_{\rm side}$ are the numbers of 
events in sideband regions for data and MC, respectively, $b_0$ is the 
expected number of background events, $s$ is the observed number of 
signal events and $s_0$ is the upper limit on the number of signal events. 
}
 \label{tab_af_open}
 \footnotesize
   \begin{tabular}{|l|l||c|c|c|c|c|c|c|}
 \hline
 \hline
 Mode    & Subdecay  & ~~~~$\epsilon$~~~~~ &  ${\mathcal{B}}_{M^0}$ & ~$N^{\rm{DATA}}_{\rm{side}}$~ & ~$N^{\rm{MC}}_{\rm{side}}$~ & $b_0$ & ~~~~$s$~~~~ & ~~~$s_0$~~~ \\
         &           mode &            (\%) &             (\%) &                             (ev.)&                            (ev.)&                              (ev.)
  &       (ev.) &       (ev.)  \\
 \hline  
 $\TEE$   & $\EGG$ & 5.68      & 39.43                         & 2   &   0             & $0.23 \pm 0.16 $      &  0  & 2.3   \\
          & $\E3P$ & 6.84      & 22.6                          & 1   &   0             & $0.23 \pm 0.23 $      &  0  & 2.2   \\
 $\TME$   & $\EGG$ & 8.03      & 39.43                         & 9   &  $9.2 \pm 2.3$  & $ 3.9   \pm 1.5 $     &  1  & 1.4   \\
          & $\E3P$  & 7.15     & 22.6                          & 2   &  $0.8 \pm 0.3$  & $0.60 \pm 0.42 $      &  0  & 1.9   \\
 $\TEP$   & $\pi^0 \RA \gamma\gamma$ & 4.70   & 98.798         & 0   & $0.7 \pm 0.7$   & $0.0^{+0.4}_{-0.0} $  &  0  & 2.4   \\
 $\TMP$   & $\pi^0 \RA \gamma\gamma$ & 6.36   & 98.798         & 16  & $12.5 \pm 2.7$  & $ 3.0 \pm 0.9      $  &  5  & 6.9   \\
 $\TEEP$  & $\EP$ & 8.51      & 17.5                           & 2   & $0.8\pm 0.3$    & $0.28 \pm 0.20 $      &  1  & 4.2   \\
 $\TMEP$  & $\EP$ & 8.41      & 17.5                           & 5   & $5.5\pm 1.9$    & $ 0.94 \pm 0.60   $   &  0  & 1.6   \\
 \hline\hline
 \end{tabular}
\end{table}

\begin{figure}[p]
\centerline{
 \resizebox{0.34\textwidth}{0.31\textwidth}{\includegraphics{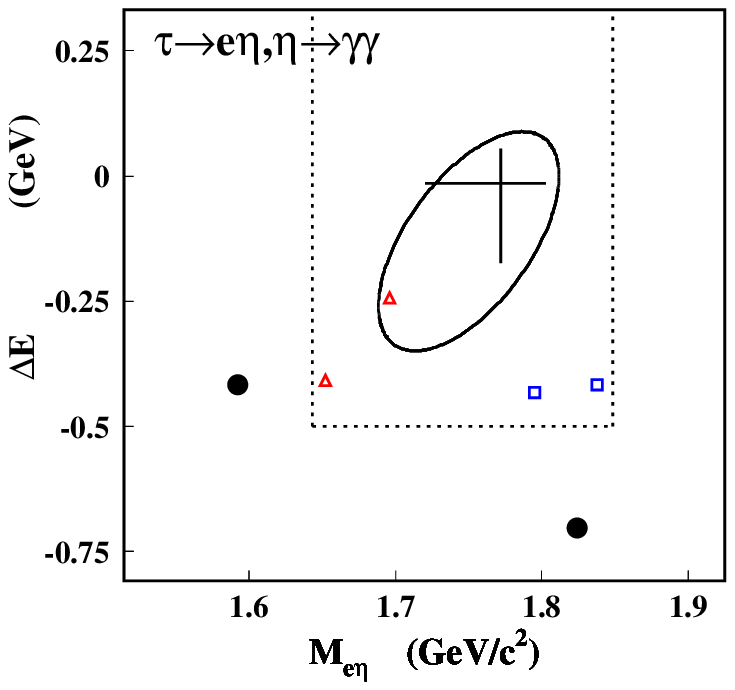}} 
 \resizebox{0.34\textwidth}{0.31\textwidth}{\includegraphics{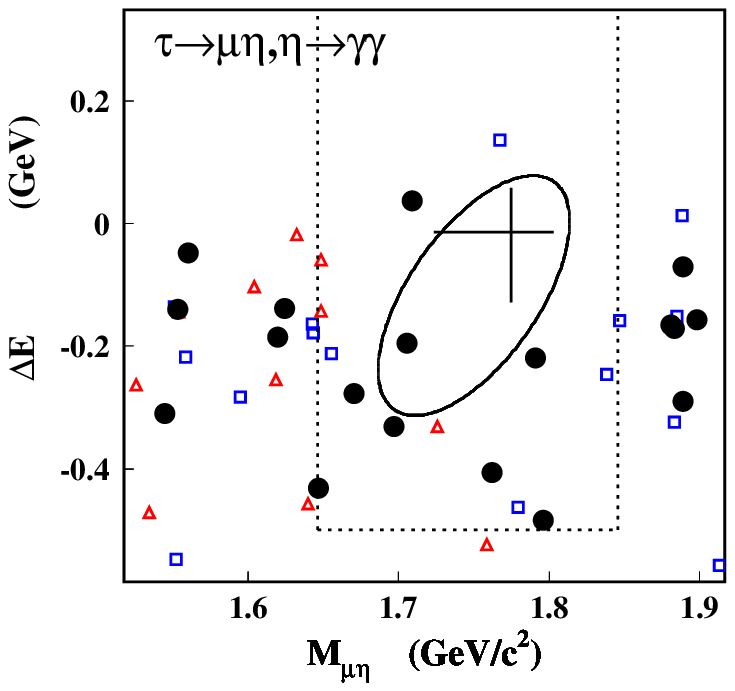}} 
}
\centerline{	  
 \resizebox{0.34\textwidth}{0.31\textwidth}{\includegraphics{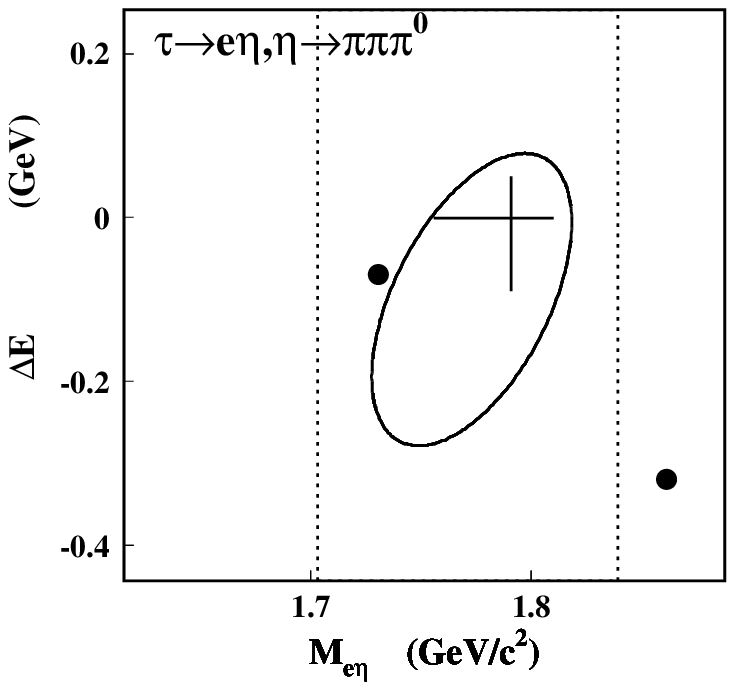}} 
 \resizebox{0.34\textwidth}{0.31\textwidth}{\includegraphics{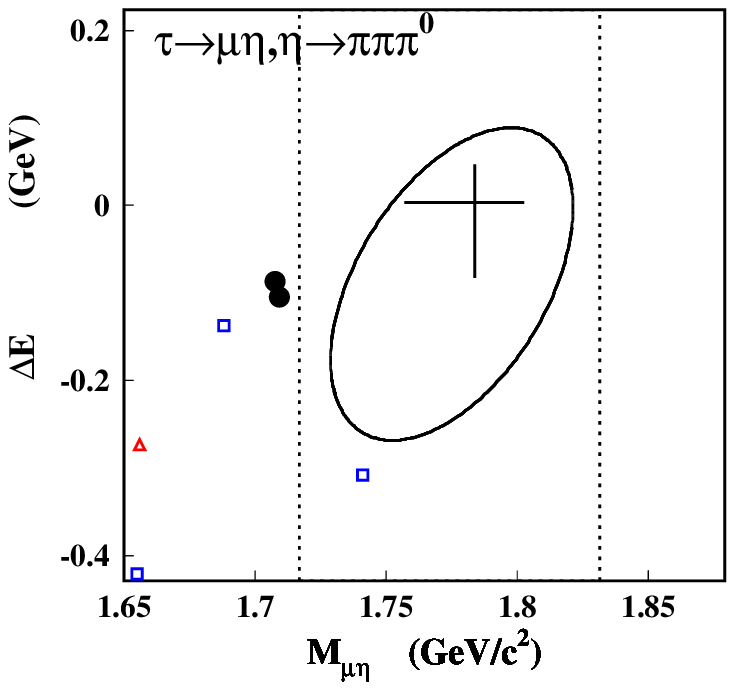}} 
}
\centerline{
 \resizebox{0.34\textwidth}{0.31\textwidth}{\includegraphics{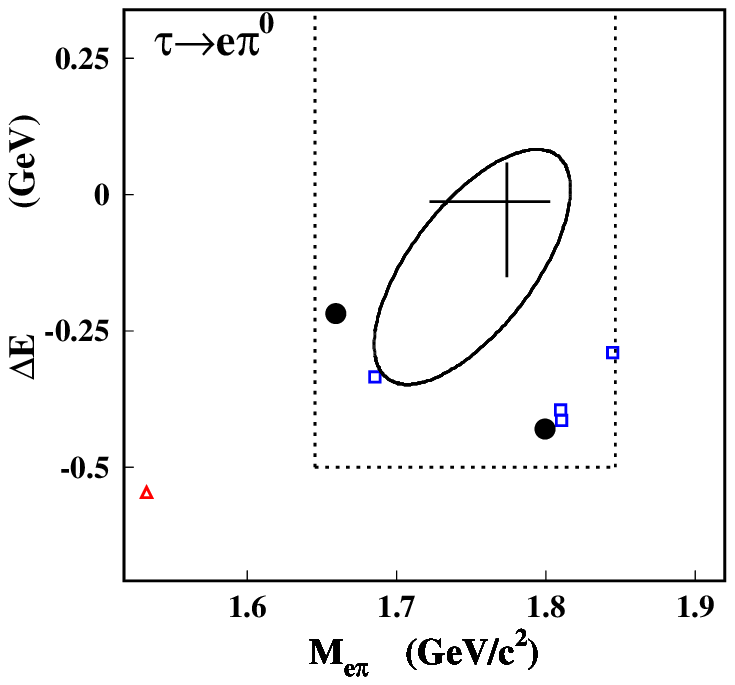}} 
 \resizebox{0.34\textwidth}{0.31\textwidth}{\includegraphics{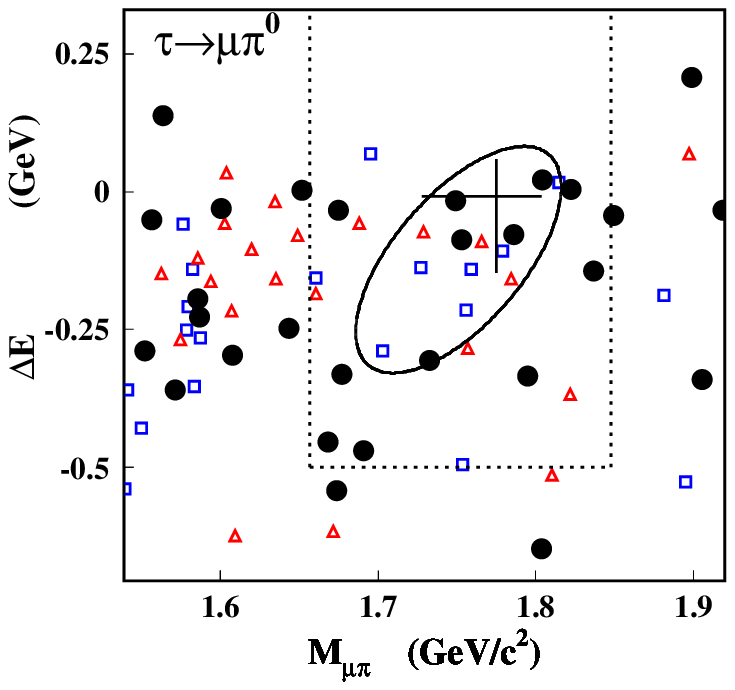}} 
}		  
\centerline{
 \resizebox{0.34\textwidth}{0.31\textwidth}{\includegraphics{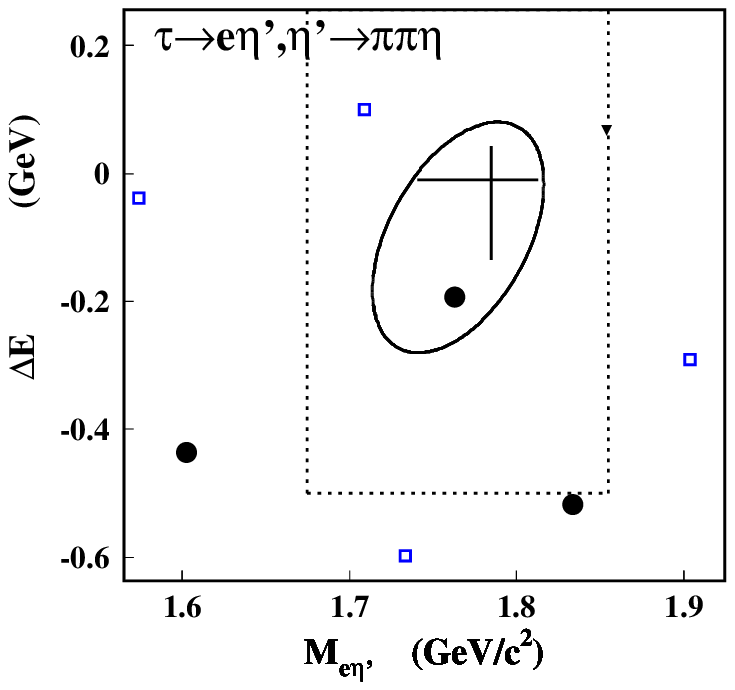}} 
 \resizebox{0.34\textwidth}{0.31\textwidth}{\includegraphics{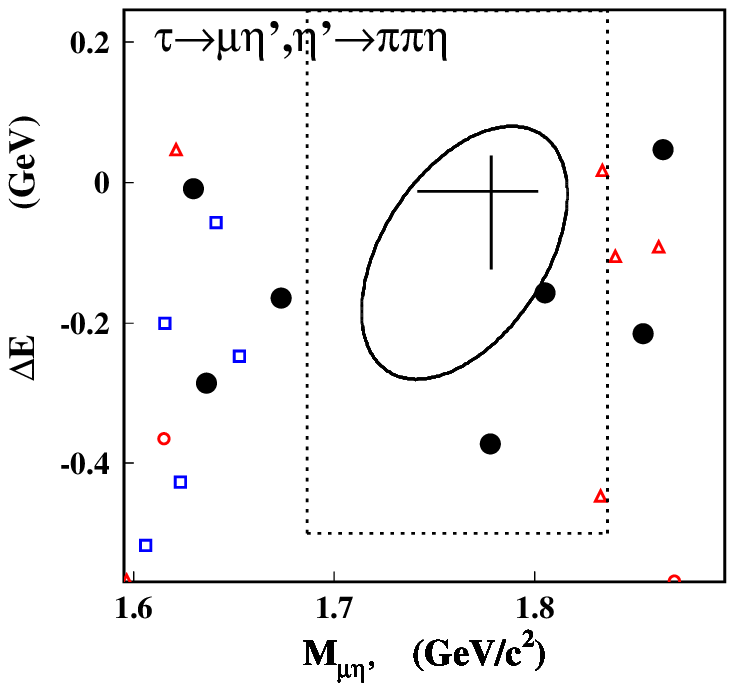}} 
}
\caption{ \small 
The distribution of remaining events over the
$\pm 10 \sigma $ region in the $ \Minv - \Delta E $
plane
after opening the blinded region. 
The ellipse is the signal region that has a signal acceptance of $\Omega = 90\%$.
The blinded regions $\pm 5\sigma$ in $\Minv$ 
and $\pm 0.5$ GeV in $\Delta E$ are indicated by the
dotted lines. The cross indicates the 
$\pm 2\sigma$ $\Minv$ and $\Delta E$ ranges. 
Various
symbols show events from
428 fb$^{-1}$ of generic $\tau^+\tau^-$ MC (\textcolor{blue}{$\Box$}), 
232 fb$^{-1}$ of $q\bar{q}$ MC with $q = u,d,s$ (\textcolor{red}{$\vartriangle$}), 
160 fb$^{-1}$ with $q = c$ (\textcolor{red}{$\circ$}), 
205 fb$^{-1}$ of two-photon MC (\textcolor{black}{$\blacktriangledown$}), 
and data (\textcolor{black}{$\bullet$}).
}
\label{fig_af_open}
\end{figure}

After opening the blinded region,
we find one event in both the $\TEEP$ and  $\TME$ modes, 
and five events in the $\TMP$ mode, 
however, no events are found for the other modes
in the elliptical signal region.  
Figure~\ref{fig_af_open} 
shows the $M_{\ell^- M^0}$--$\Delta E$ plot for 
the individual modes. 
The resultant numbers of events $s$ are compared to 
the expected
background
$b_0$ in Table~\ref{tab_af_open}:
good agreement
can be observed.

Following the Feldman-Cousins method~\cite{FC}, 
we calculate the upper limit $s_0$ on the number of signal events
at the 90\% confidence level (C.L.) for all modes, 
as listed in Table \ref{tab_af_open}.
These values take only statistical errors into account.

The branching fraction at the 90\% C.L. is obtained from
the following formula, 
\begin{equation}
{\mathcal{B}}(\tau^- \RA \ell^- M^0) = \frac{s_0}{2~\epsilon~{\mathcal{B}}_{M^0} N_{\TT}},  
\label{eq4}
\end{equation}
where
$\epsilon$ is the detection 
efficiency for the individual modes,  
${\mathcal{B}}_{M^0}$ is the branching fraction for the 
$M^0$ decay,
and 
$N_{\tau\tau} = 137.2\times 10^{6}$ is the number of 
produced $\tau$-pairs at 153.8 fb$^{-1}$ luminosity, 
with the cross section 
of $e^+e^- \rightarrow \tau^+\tau^-$ at the $\Upsilon(4S)$ resonance
$\sigma_{\tau\tau} = 0.892\pm 0.002$ nb  
calculated by KKMC~\cite{kkmc}.
The values of $\epsilon$ and $\mathcal{B}_{M^0}$ 
for each mode
are listed in 
Table~\ref{tab_af_open}.
Systematic uncertainties due to the background estimate
$b_0$, and uncertainties on $2 \epsilon \mathcal{B}_{M_0} N_{\tau \tau}$,
are taken into account by inflating the value of $s_0$, as
discussed below.

\begin{table}[t]
 \caption{\small Systematic uncertainties in \%. }
 \label{tab_sys}
 \footnotesize
\begin{tabular}{|l||c|c|c|c|c|c|c|c|}
\hline\hline
Mode~~~$\tau^- \RA$ & $\MTEE, $ & $\MTEE, $ &   $\MTME, $ & $\MTME, $ &
$\MTEP$ & $\MTMP$ & $\MTEEP$ & $\MTMEP$ \\
& $\EGG$ & $\EIP3$ & $\EGG$ & $\EIP3$ & & & &  \\
\hline
Track recon.                     & 2.0 & 2.0 & 2.0 & 2.0 & 2.0 & 2.0 & 2.0 & 2.0 \\
$M^0$ recon.                     & 2.0 & 4.0 & 2.0 & 4.0 & 2.0 & 2.0 & 4.0 & 4.0 \\
$\pi^0$ veto                     & 5.5 & --  & 5.5 & --  & --  & --  & 5.5 & 5.5 \\
$e$ ID                           & 1.0 & 1.0 & --  & --  & 1.0 & --  & 1.0 & --  \\ 
$\mu$ ID                         & --  & --  & 2.0 & 2.0 & --  & 2.0 & --  & 2.0 \\
Trigger                          & 0.5 & 0.1 & 0.2 & 0.1 & 0.7 & 0.4 & 0.1 & 0.1 \\ 
Beam background                  & 2.3 & 2.1 & 2.3 & 2.1 & 2.3 & 2.3 & 2.1 & 2.1 \\
Luminosity                       & 1.4 & 1.4 & 1.4 & 1.4 & 1.4 & 1.4 & 1.4 & 1.4 \\ 
$\mathcal{B}_{M^0}$              & 0.7 & 1.8 & 0.7 & 1.8 & --  & --  & 3.4 & 3.4 \\
MC stat.                         & 1.4 & 1.7 & 1.1 & 1.6 & 0.9 & 0.8 & 1.2 & 1.1 \\
MC models                        & 0.5 & 0.5 & 0.5 & 0.5 & 0.5 & 0.5 & 0.5 & 0.5 \\
\hline         
Total                            & 7.0 & 5.8 & 7.2 & 6.0 & 4.2 & 4.5 & 8.4 & 8.6 \\
\hline \hline
\end{tabular}
\end{table}

The systematic uncertainties on the detection sensitivity,
$2\epsilon{\mathcal{B}}_{M^0} N_{\TT}$, 
arise from the track reconstruction efficiency
(1.0\% for each track)
of the tag side track and the signal side lepton;  
$\pi^0$, $\eta$ and $\eta'$ reconstruction efficiency 
which includes the uncertainties
in the tracking efficiency for charged pions (2.0\%)
and the $\pi^0$ or $\eta$ reconstruction efficiency (2.0\%);
$\pi^0$ veto for $\tau^- \RA \ell^- \eta$ and $\ell^- \eta'$ modes (5.5\%); 
electron or muon identification efficiency (1.0\% for electron, 2.0\% for muon); 
trigger efficiency (0.1--0.7\%);
beam background effect (2.1\% for 1-1 prong, 2.3\% for 1-3 prong events); 
luminosity (1.4\%); 
branching fraction of $\pi^0$, $\eta$, and $\eta'$ (Ref.~\cite{pdg});
signal MC statistics (0.8--1.7\%)
and signal MC models (0.5\%).
Table ~\ref{tab_sys} lists these separate contributions
as well as the resulting systematic 
uncertainties obtained by adding all the components in quadrature.
The dominant contributions, 
$\pi^0$, $\eta$ and $\eta'$ reconstruction efficiencies, 
$\pi^0$ veto efficiency
and
beam
background
effect, 
are estimated as follows.
The contribution of reconstruction efficiencies for pseudoscalar mesons
is evaluated from the efficiency ratios for data and MC samples
using $\eta \rightarrow \gamma\gamma$ and $\eta \rightarrow 3\pi^0$ decays.
The $\pi^0$ veto contribution is
also evaluated by the efficiency ratio of data and MC,
where the $\pi^0$ veto efficiency is 
estimated from the difference of $\eta$ reconstruction efficiencies
with and without the $\pi^0$ veto.
The beam
background
effect is estimated
from
data that is taken by the random trigger
at the same time
as
the normal data taking.
We initially assumed
a uniform angular distribution
of $\tau$ decays for the signal MC.
Its possible nonuniformity
was taken into account by comparing the uniform case 
with 
those
assuming $V-A$ and $V+A$ interactions~\cite{spin}, 
which result in maximum possible variations.
This effect contributes 0.5\% shown in the
``MC models''
line of Table~\ref{tab_sys}.

The treatment of systematic uncertainties depends on the
estimated background $b_0$. In cases where $b_0 < 1.0$, 
except for $\tau^- \rightarrow \mu^- \eta'$, we set $b_0$ to $0.0\pm 0.0$
to obtain a conservative upper limit. 
We use the POLE program~\cite{pole}
to recalculate $s_0$ including
systematic uncertainties on both $b_0$ and the detection
sensitivity, assuming Gaussian errors~\cite{Cous}. The upper 
limits then obtained from equation (\ref{eq4}) are summarized in
Table~\ref{tab_UL}.

\begin{table}[h]
 \caption{\small Upper limits on branching fractions. }
 \label{tab_UL}
 \begin{center}
 \small
   \begin{tabular}{|l|l||c|}
 \hline
 \hline
 Mode    & Subdecay mode & U.L. of ${\mathcal{B}}$ @ 90\% C.L. \\
 \hline  
 $\TEE$   & $\EGG$   & $ 4.0   \times 10^{-7}$ \\
          & $\E3P$   & $ 5.8   \times 10^{-7}$ \\
 $\TEE$   & combined  & $ 2.4   \times 10^{-7}$ \\
\hline		    
 $\TME$   & $\EGG$   & $ 2.3   \times 10^{-7}$ \\
          & $\E3P$   & $ 5.5   \times 10^{-7}$ \\
 $\TME$   & combined  & $ 1.5   \times 10^{-7}$ \\
\hline
 $\TEP$   & $\pi^0 \RA \gamma\gamma$ & $ 1.9   \times 10^{-7}$ \\
 $\TMP$   & $\pi^0 \RA \gamma\gamma$ & $ 4.1   \times 10^{-7}$ \\
\hline
 $\TEEP$  & $\EP$ & $ 10    \times 10^{-7}$ \\
 $\TMEP$  & $\EP$ & $ 4.7   \times 10^{-7}$ \\
 \hline\hline
 \end{tabular}
 \end{center}
\end{table}

\section{Discussion}

One can see from Table \ref{tab_UL} that
for the $\TEEP$ and $\TMEP$ modes, 
the resulting 90\% C.L. upper limits are
${\mathcal{B}}(\TEEP) < 10 \times 10^{-7}$
and
${\mathcal{B}}(\TMEP) < 4.7 \times 10^{-7}$;
these are the first experimental limits on these modes.
For the $\TEE$ and $\TME$ modes, the results from two different 
final states, $\eta\RA\gamma\gamma$ and $\eta\RA\pi^+\pi^-\pi^0$, are combined. 
The resulting 90\% C.L. upper limits 
for these two modes and 
$\tau^- \rightarrow e^- \pi^0$
and
$\tau^- \rightarrow \mu^- \pi^0$
are ${\mathcal{B}}(\TEE) < 2.4 \times 10^{-7}$, 
${\mathcal{B}}(\TME) < 1.5 \times 10^{-7}$, 
${\mathcal{B}}(\TEP) < 1.9 \times 10^{-7}$ and 
${\mathcal{B}}(\TMP) < 4.1 \times 10^{-7}$ 
and improve upon the previous CLEO measurements
by a factor of
34, 64, 20 and 10, 
respectively.

\begin{figure}[h]
\centerline{
 \epsfxsize=9.5cm  \epsfbox{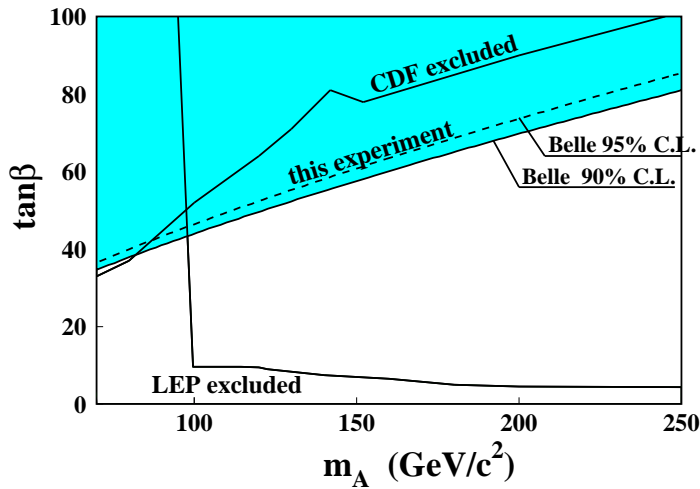}
}
\caption{\small Experimentally excluded $m_A-\tan\beta$ parameter space. 
The 90\% C.L. result of this experiment on 
$\mathcal{B}(\tau^- \rightarrow \mu^-\eta)$
using the relation (\ref{eq:Sher})
is
indicated by the shaded region
together with the 95\% C.L. regions excluded by 
this experiment
(dashed line), 
LEP~\cite{LEP} and the CDF~\cite{Tevatron,pdg}.}
\label{tanb_ma}
\end{figure}

We can restrict the allowed parameter space for $m_A$--$\tan\beta$,
using the relationship derived by M.~Sher~\cite{Sher} 
in a seesaw MSSM with a specific mass texture:
\begin{equation}
  {\mathcal{B}}(\tau^- \!\! \rightarrow\mu^{\!-}\eta)= 
 0.84\times 10^{-6} \! \times \! 
 \left(\frac{\rm{tan}\beta}{60}\right)^{6} 
 \left(\frac{100~\rm{GeV}}{m_A}\right)^{4}, 
 \label{eq:Sher}
\end{equation}
where $m_A$ is the pseudoscalar Higgs mass and $\rm{tan}\beta$ is the ratio 
of the vacuum expectation values $(\langle H_u\rangle/\langle H_d\rangle)$, 
as shown in Fig.~\ref{tanb_ma}, 
where our boundary is indicated 
for the 90\% C.L by the shaded region. 
Figure~\ref{tanb_ma}
also shows the 95\% C.L. constraints
from
our experiment with
a
dashed line and 
high energy collider experiments at LEP~\cite{LEP} and 
CDF~\cite{Tevatron}. 
Our experiment has a sensitivity competitive with that of the CDF experiment 
which searched for 
$pp\rightarrow A/\phi b\overline{b}\rightarrow b\overline{b}b\overline{b}$ reactions, 
where $\phi$ is a CP-even neutral Higgs state and $A$ is a CP-odd state 
in the MSSM. 

The improved sensitivity to rare $\tau$ lepton decays achieved in this 
work can also be used to constrain the parameters of models with heavy 
Dirac neutrinos~\cite{gonzalez,ilakovac}. In these models the expected branching 
ratios of various LFV decays are evaluated in terms of combinations of 
the model parameters.  These combinations, denoted $y_{\tau e}$ and 
$y_{\tau \mu}$ for $\tau$ decays involving an electron and a muon,
respectively, can vary from 0 to 1. Our $\tau^- \to e^- \pi^0$ result
sets a 90\% C.L. upper limit $y_{\tau e} < 0.26$, the most restrictive
bound on this quantity. The corresponding limit from our 
$\tau^- \to \mu^-\pi^0$ result is $y_{\tau \mu} < 0.87$: a somewhat better
bound $y_{\tau \mu} < 0.65$ can be set from the 
$\tau^- \to \mu^- \mu^+ \mu^-$ decay based 
on the experimental limits from BaBar~\cite{babar_3l} and Belle~\cite{belle_3l}.

In summary, 
using a data sample of 153.8 fb$^{-1}$ collected with the Belle detector
we obtained new upper limits on the branching fractions
of semileptonic LFV $\tau^-$ decays
involving pseudoscalar mesons $\pi^0,~\eta$ and $\eta'$. 
They range from 
$1.5 \times 10^{-7}$
to $10  \times 10^{-7}$
for the six decay modes
studied and are 10 to 64 times more restrictive
than previous limits.

\section{Acknowledgements}

We thank the KEKB group for the excellent
operation of
the accelerator, the KEK cryogenics group
for the efficient operation of the solenoid, 
and the KEK computer group and the National Institute of Informatics
for valuable computing and Super-SINET network support. 
We are grateful to A.~Ilakovac for fruitful discussions.
We acknowledge support from the Ministry of Education, 
Culture, Sports, Science, and Technology of Japan
and the Japan Society for the Promotion of Science;
the Australian Research Council
and the Australian Department of Education, Science and Training;
the National Science Foundation of China under contract No.~10175071;
the Department of Science and Technology of India;
the BK21 program of the Ministry of Education of Korea
and the CHEP SRC program of the Korea Science and Engineering Foundation;
the Polish State Committee for Scientific Research
under contract No.~2P03B 01324;
the Ministry of Science and Technology of the Russian Federation;
the Ministry of Higher Education, Science and Technology of the Republic of Slovenia;
the Swiss National Science Foundation;
the National Science Council and the Ministry of Education of Taiwan;
and the U.S.\ Department of Energy.

\end{document}